\documentclass[conference]{IEEEtran}
\usepackage[a-1b]{pdfx}
\PassOptionsToPackage{bookmarks=false,hidelinks}{hyperref}

\usepackage[utf8]{inputenc}
\usepackage[T1]{fontenc}
\usepackage{amsmath,amssymb}
\usepackage{graphicx}
\usepackage{array}
\usepackage{tabularx}
\usepackage{cite}
\usepackage{xcolor}
\usepackage{url}

\newcolumntype{L}{>{\raggedright\arraybackslash}X}
\newcolumntype{C}{>{\centering\arraybackslash}X}

\newcommand{\figimage}[2]{%
  \IfFileExists{#1.png}{\includegraphics[width=\figwidth]{#1.png}}{%
  \IfFileExists{#1.pdf}{\includegraphics[width=\figwidth]{#1.pdf}}{%
  \IfFileExists{#1.jpg}{\includegraphics[width=\figwidth]{#1.jpg}}{%
  \fbox{\parbox[c][\dimexpr#2\dimexpr\figwidth\relax\relax][c]{\dimexpr\figwidth-2\fboxsep-2\fboxrule\relax}{\centering\footnotesize
    PLACEHOLDER: replace with \texttt{#1}}}}}}}

\AtBeginDocument{\def\IEEEbibitemsep{0pt plus 0.3pt}}
\newcommand{\figwidth}{0.72\columnwidth}

\usepackage{fancyhdr}

\fancypagestyle{firstpage}{
    \fancyhf{}
    \fancyhead[C]{%
        \parbox{\textwidth}{%
            \centering
            \scriptsize
            © 2026 IEEE. Personal use of this material is permitted.
            Permission from IEEE must be obtained for all other uses, in any
            current or future media, including reprinting/republishing this
            material for advertising or promotional purposes, creating new
            collective works, for resale or redistribution to servers or lists,
            or reuse of any copyrighted component of this work.
        }%
    }
    
}

\begin{document}

\title{LLMs for Software Vulnerability Prediction and Prioritization: A Review}

\author{
\parbox[t]{0.49\textwidth}{\centering
Faheem Ullah\\
\mbox{\textit{College of Interdisciplinary Studies}}\\
\mbox{\textit{Zayed University}}\\
Abu Dhabi, UAE\\
faheem.ullah@zu.ac.ae}
\hfill
\parbox[t]{0.49\textwidth}{\centering
Muhammad Moiz Khalid\\
\mbox{\textit{Department of Artificial Intelligence and Data Science}}\\
\mbox{\textit{National University of Computer and Emerging Sciences}}\\
Islamabad, Pakistan\\
i232552@isb.nu.edu.pk}
\\[2ex]
\parbox[t]{0.49\textwidth}{\centering
Shayan Ayaz Khan\\
\mbox{\textit{School of Science and Engineering}}\\
\mbox{\textit{Lahore University of Management Sciences}}\\
Lahore, Pakistan\\
27100290@lums.edu.pk}
\hfill
\parbox[t]{0.49\textwidth}{\centering
Hussain Ahmad\\
\mbox{\textit{School of Computer Science}}\\
\mbox{\textit{The University of Adelaide}}\\
Adelaide, Australia\\
hussain.ahmad@adelaide.edu.au}
}

\maketitle
\thispagestyle{firstpage}
\begin{abstract}
Large language models have moved beyond vulnerability detection into prediction, assessment, exploitability estimation, risk assessment and prioritization, yet the evidence for this shift has not been synthesised separately from the detection literature. In this systematic literature review, we examine how LLMs are used across those five task classes, how well they perform against established baselines, and what prevents their operational adoption. Following a prespecified protocol, we identified 197 records, removed 53 duplicates, and excluded 103 records at title and abstract stage. We assessed 41 full texts and finally retained 23 primary studies published between 2024 and 2026 after a six-question quality assessment. The review contributes a five-class taxonomy ordered by decision outcome, a method taxonomy covering prompting, tuning and retrieval or agentic architectures, an evaluation-practice audit, a bounded performance synthesis, and a deployment-barrier analysis. The main findings are that (i) assessment is the largest class (9 of 23 studies), ahead of risk assessment and prioritization (7 each); (ii) reported accuracy is highest where the label space is tightest and is not stable across evaluation periods; (iii) explanation usability is an audited evidence gap rather than a demonstrated capability; and (iv) reproducibility is the weakest reported dimension. We conclude that present systems should support analyst judgement rather than replace it, and we specify three validation studies that would settle the open questions.
\end{abstract}

\begin{IEEEkeywords}
large language models, software vulnerability, vulnerability prioritization, CVSS, exploitability prediction, risk assessment, systematic literature review.
\end{IEEEkeywords}

%% ====================================================================
\section{Introduction}\label{sec:intro}

\begin{figure}[!t]
\centering
% PLACEHOLDER FOR FIGURE 2: replace figure2.png (originally fig1_year_trend.png)
\figimage{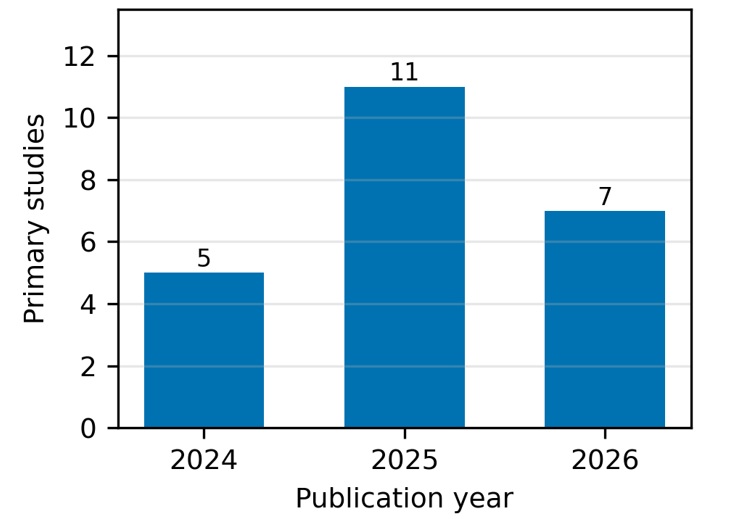}{0.80}
\caption{Primary-study distribution by year of publication.}
\label{fig:year}
\end{figure}

This systematic literature review maps how large language models are used for software vulnerability prediction, assessment, exploitability estimation, risk assessment and prioritization, deliberately excluding detection-only work as the primary scope. We synthesised twenty-three primary studies published between 2024 and 2026 (Fig.~\ref{fig:year}) from an initial pool of 197 records identified across our search sources, and retained eighteen secondary studies only for framing. The analysis that follows cites and discusses every one of the twenty-three primary studies. The principal finding is a taxonomy that orders the field by outcome rather than by technique: prediction, assessment, exploitability prediction, risk assessment and prioritization form a chain downstream of detection, and each class commits the analyst to a different kind of decision. Assessment is the largest class, with risk assessment and prioritization following at seven studies each, so the class closest to operational remediation decisions is smaller and less mature than assessment in this corpus. Methodologically the systems are more uniform than the tasks: most begin from a general-purpose chat model and add task-specific machinery at one of three levels, namely crafted prompting, task-specific tuning, or retrieval and multi-agent orchestration that supplies external context before a judgement is produced. 

Evaluation practice is the weak point. Ground truth is concentrated on public vulnerability records with CVSS, CWE, KEV and EPSS labels, but studies do not standardise label definitions and especially train-test splits, and random splits drawn from a single period cannot establish prospective validity given documented contamination and knowledge-cutoff effects. Reported accuracy is consequently high for severity and scoring tasks and competitive but qualified for exploitability and prioritization, and the advantage over classical baselines is not stable across evaluation periods. Unverified explanation usability, which no included study tests with human analysts, weak reproducibility of prompts and orchestration code, and the cost and privacy friction of sending sensitive advisory and code context to hosted models further constrain deployment. The practical implication is that analysts should treat current systems as evidence-gathering \cite{ahmad2025immersive} aids rather than as autonomous scorers or rankers, and that the highest-value next step is a time-split, cost-reported, analyst-validated evaluation of a prioritization pipeline.

%% ====================================================================
\section{Research Methodology}\label{sec:method}

This review follows our prespecified protocol and reports its conduct against PRISMA-style stages. The unit of analysis is the primary study; we exclude detection-only work and secondary surveys from synthesis but retain them for background and for snowballing. The protocol prespecified the scope boundary, the search string, the publication window of January 2020 to September 2026, seven inclusion criteria and eight exclusion criteria (Table~\ref{tab:criteria}), a six-question quality instrument, a data-extraction schema (Table~\ref{tab:extraction}) and a five-class taxonomy. Applying those rules to the retrieved corpus yields the study set analysed in the following sections, and applying them before we inspected results keeps the inclusion threshold independent of the findings.

\subsection{Protocol, Scope Boundary and Research Questions}

The organising boundary is the distinction between detection, prediction, assessment and prioritization. Detection answers whether a vulnerability exists in a given artefact; prediction answers whether a component is likely to contain or to develop vulnerabilities; assessment answers how severe, risky or exploitable a known vulnerability is; prioritization answers which vulnerabilities should be addressed first. Only the last three families, together with exploitability and risk estimation, constitute the primary scope here. This boundary is what separates the present review from the large body of LLM vulnerability-detection surveys, and it is also the first analytical result, because several retrieved records describe detection systems whose abstracts contain no prediction, scoring or ranking outcome, and we therefore excluded them under EC1. The five research questions map directly onto that boundary: what LLMs do, how they are built, what they are evaluated on, how well they perform, and what prevents deployment.

\subsection{Search Strategy, Databases and Screening}

The master search string combined three concept blocks: a language-model block, a software-vulnerability block, and an outcome block covering predict, priorit, rank, assess, sever, exploit, risk, scoring, classification and forecast variants. We reserved a narrower variant for sources returning excessive irrelevant material, and we record the exact query issued against each source rather than collapse it into a single string, because search syntax and field coverage differ across platforms. The protocol named six bibliographic databases (IEEE Xplore, the ACM Digital Library, Scopus, Web of Science, ScienceDirect and SpringerLink), with Google Scholar as a supplementary source. In this review, direct subscription access to those six platforms was not available, so retrieval ran through OpenAlex, which indexes the venues those databases cover, together with arXiv for preprints, Semantic Scholar where rate limits permitted, DOI-level full-text fetching, and the official CISA, FIRST and MITRE registries for scoring and exploitation ground truth. As OpenAlex lacks the native search features of the original databases, coverage may be incomplete, which we partly mitigated through DOI-level full-text retrieval and citation chaining. This substitution is a documented deviation from the protocol, and we revisit it among the threats to validity (Table~\ref{tab:threats}).

\subsection{PRISMA Selection Outcomes and Quality Assessment}

Screening proceeded from identification through deduplication, title and abstract screening, full-text assessment and quality scoring. Table~\ref{tab:sources} lists the records pooled by source, and the selection funnel was as follows: we identified 197 records, removed 53 duplicates, excluded 103 records at title and abstract stage, assessed 41 full texts and removed 18 secondary studies from the primary set under EC8, leaving 23 primary studies in the final synthesis. We still use retained secondary studies for framing and citation chaining. Five of the 23 are boundary studies, retained as scope exceptions to IC3 under IC4 because they provide substantial technical analysis bearing on the evaluation and validity of these systems: two threat-intelligence benchmarks \cite{alam2024ctibench,cheng2026cticonnect}, a contamination survey \cite{xu2024contamination}, and two repair studies with exploit-based verification \cite{wang2025vulnrepaireval,seo2025autopatch}. They support RQ3 and RQ5; two of them \cite{wang2025vulnrepaireval,seo2025autopatch} are still counted in Fig.~\ref{fig:classes} under their closest class, and the survey is an exception to EC8 because it is the only study characterising contamination. Eight of the 23 are not peer-reviewed as listed: seven arXiv preprints \cite{shen2025vulstamp,foundjem2025multiagent,tang2025polar,wang2025vulnrepaireval,alam2024ctibench,xu2024contamination,seo2025autopatch} and one Zenodo deposit \cite{dhok2026mcp}. We scored quality against six questions covering objective clarity, method description, dataset description, metric and baseline suitability, result reporting, and discussion of limitations, with Yes scoring one, Partially one half and No zero. Scores cluster at the middle of the range, because studies usually describe dataset provenance at registry level rather than at row level and because limitations and reproducibility are the most frequently neglected dimensions; the distribution appears in Table~\ref{tab:quality} and we revisit it under RQ5.

% == Table I ==
\begin{table*}[!t]
\centering
\caption{Search Sources, Queries and Execution Status}
\label{tab:sources}
\scriptsize\renewcommand{\arraystretch}{1.0}
\begin{tabularx}{\textwidth}{L L r}
\hline
\textbf{Source / lane} & \textbf{Query concept blocks} & \textbf{Records pooled}\\
\hline
OpenAlex (indexes IEEE, ACM, Springer, Elsevier, Scopus/WoS venues) & LM block AND vulnerability block AND outcome block; 11 query cells & $\sim$145\\
arXiv (preprints) & same master string, preprint filter & 15\\
Semantic Scholar (supplementary) & same master string (we reran rate-limited cells elsewhere) & 10\\
DOI / publisher full text & targeted lookups of candidate records & 23\\
CISA KEV, FIRST EPSS, FIRST CVSS v4.0, MITRE CWE (official registries) & scoring, exploitation and weakness ground truth & 4\\
\hline
Total pooled before deduplication & & 197\\
\hline
\end{tabularx}
\end{table*}

% == Table II ==
\begin{table*}[!t]
\centering
\caption{Inclusion and Exclusion Criteria Applied During Screening}
\label{tab:criteria}
\scriptsize
\renewcommand{\arraystretch}{1.0}
\begin{minipage}[t]{0.495\textwidth}
\begin{tabularx}{\linewidth}{@{}l L L@{}}
\hline
\textbf{ID} & \textbf{Criterion} & \textbf{Decision}\\
\hline
IC1 & Investigates LLMs or generative AI & include if IC2-IC7 also hold\\
IC2 & Concerns software vulnerabilities or weaknesses & include\\
IC3 & Addresses prediction, assessment, scoring, exploitability, severity, ranking or prioritization & include\\
IC4 & Provides empirical evaluation, method, framework or substantial technical analysis & include\\
IC5 & Full text (or substantive indexed abstract record) available & include\\
IC6 & Written in English & include\\
IC7 & Peer-reviewed journal, conference or workshop paper, or credible scholarly preprint or deposit with substantial technical detail & include\\
\hline
\end{tabularx}
\end{minipage}\hfill
\begin{minipage}[t]{0.495\textwidth}
\begin{tabularx}{\linewidth}{@{}l L L@{}}
\hline
\textbf{ID} & \textbf{Criterion} & \textbf{Decision}\\
\hline
EC1 & Pure detection with no prediction/risk/prioritization component & exclude\\
EC2 & Pure repair or patch generation & exclude\\
EC3 & LLM self-security (prompt injection, jailbreaking) without vulnerability prioritization relevance & exclude\\
EC4 & Traditional ML/DL with no LLM component & exclude\\
EC5 & Blog, tutorial, editorial or poster without technical evidence & exclude\\
EC6 & Non-English publication & exclude\\
EC7 & Duplicate publication & exclude\\
EC8 & Secondary study (SLR or survey) & exclude from primary synthesis; retain for background\\
\hline
\end{tabularx}
\end{minipage}
\end{table*}

% == Table III ==
\begin{table}[!t]
\centering
\caption{Quality Assessment Against QA1-QA6 (23 Primary Studies)}
\label{tab:quality}
\scriptsize\renewcommand{\arraystretch}{1.0}
\begin{tabularx}{\columnwidth}{@{}L >{\centering\arraybackslash}p{0.11\columnwidth} >{\centering\arraybackslash}p{0.15\columnwidth} >{\centering\arraybackslash}p{0.17\columnwidth}@{}}
\hline
\textbf{QA item} & \textbf{Mean (0-1)} & \textbf{Studies scoring 1} & \textbf{Studies scoring 0.5}\\
\hline
QA1 Objectives clearly stated & 0.93 & 20 & 3\\
QA2 LLM/method adequately described & 0.78 & 13 & 10\\
QA3 Dataset/source clearly described & 0.67 & 9 & 13\\
QA4 Appropriate metrics/baselines & 0.72 & 11 & 11\\
QA5 Results clearly/reproducibly reported & 0.52 & 6 & 12\\
QA6 Limitations/threats discussed & 0.41 & 4 & 11\\
\hline
Overall (max 6) & \multicolumn{3}{c}{4.03 mean total}\\
\hline
\end{tabularx}
\end{table}

%% ====================================================================
\section{Results}\label{sec:results}

\subsection{RQ1 (What LLMs Do): A Task Taxonomy Beyond Detection}\label{sec:rq1}

The retrieved literature does not treat ``LLM for vulnerabilities'' as one task. Read by outcome rather than by title, the primary studies fall into five application classes ordered by what their output commits the analyst to: prediction commits to an expectation about future or latent weakness; assessment commits to a severity judgement about a known instance; exploitability commits to a behavioural forecast; risk commits to a consequence judgement tied to an asset; prioritization commits to an ordering that drives remediation effort.

\subsubsection{Vulnerability Prediction and Vulnerable-Component Forecasting}

The first class asks whether a component or release is likely to harbour vulnerabilities without naming a specific flaw. A structured security assessment of the Windows App SDK integrates static analysis tools behind a context-aware, disagreement-aware language-model interpretation layer, producing component-level judgements \cite{pakshad2026windows}. A related line applies language models inside probabilistic alarm-handling models that order alarms so reviewers inspect the highest-value candidates first \cite{pan2025alarms}. The predicted unit is therefore an alarm or component rather than a single flaw.

\subsubsection{Assessment: Severity, CVSS and CWE Classification}

The second class takes an identified vulnerability and estimates how bad it is, and it is the largest group in the corpus. CVSS prediction is its clearest instantiation \cite{duan2021iot}: one study compares GPT-4o-mini against BERT for predicting CVSS base vectors as a mapping from advisory text to metric values \cite{isogai2024cvss}, while a systematic assessment of automated scoring quantifies where models agree with and diverge from published scores \cite{sanvito2025autocvss}. CVE-LLM learns evaluation from historical NVD assessments assisted by an ontology, linking assessment to weakness structure \cite{ghosh2025cvellm}. Severity is not context-free: a study of vulnerability batches shows judgements shift with the product and surrounding reports \cite{miranda2025contextual}, and a fine-tuned threat-intelligence agent couples severity with contextual risk and a mitigation recommendation \cite{giriraju2026cti}. A retrieval-augmented pipeline extends the class to bug-report triage, where the label sets resolution order \cite{ali2025rag}.

\subsubsection{Exploitability, Risk and Prioritization Tasks}

\begin{figure}[!t]
\centering
% PLACEHOLDER FOR FIGURE 3: replace figure3.png (originally fig3_classes.png)
\figimage{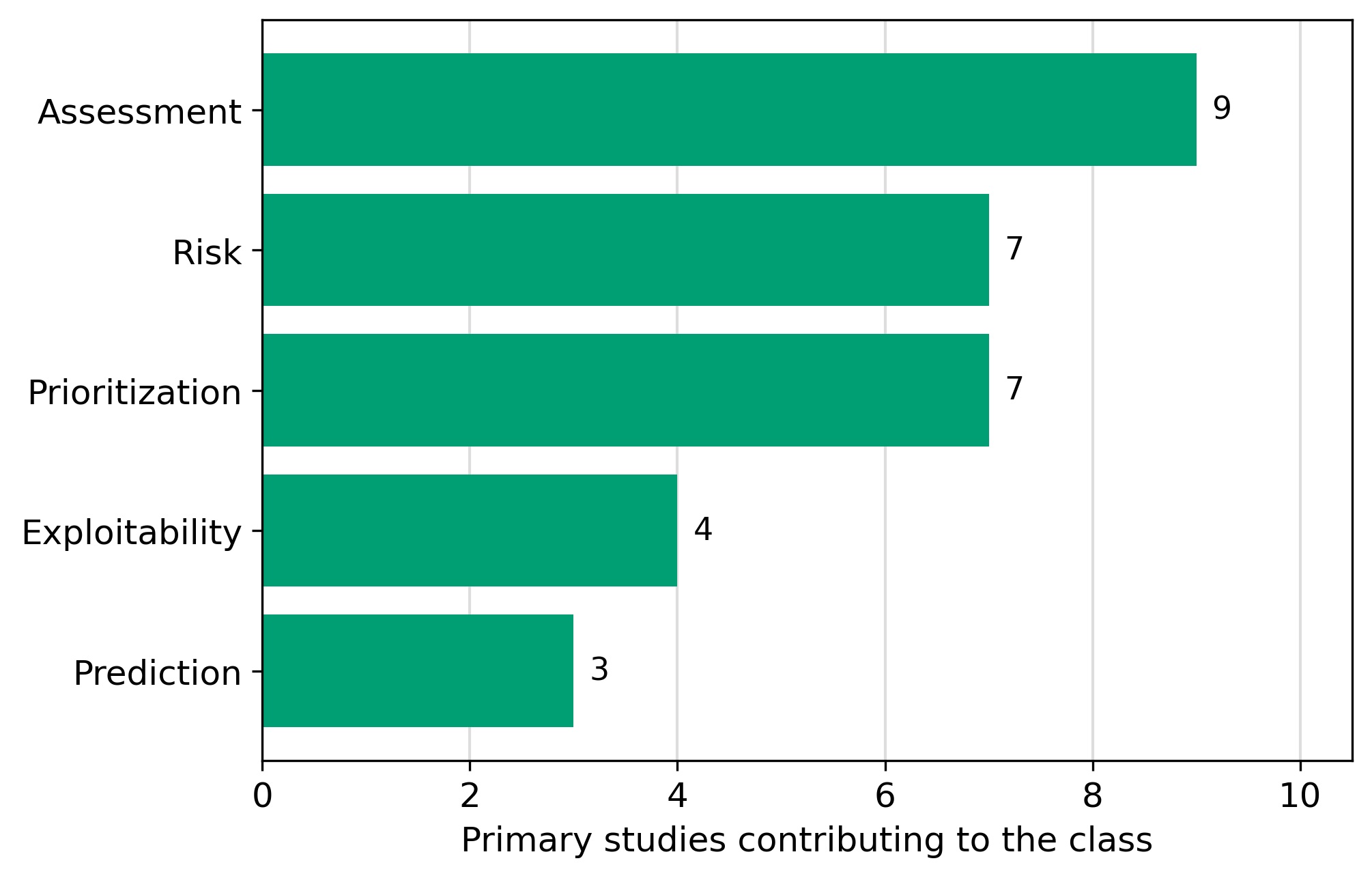}{0.69}
\caption{Distribution of primary studies across the five application classes. Excluding the two boundary studies counted here \cite{wang2025vulnrepaireval,seo2025autopatch}, assessment is 8 and prioritization 5.}
\label{fig:classes}
\end{figure}

The remaining classes forecast behaviour, judge consequence and order remediation. Exploitability studies estimate the probability of actual exploitation, motivated by the observation that most detected flaws have low exploitability and that indiscriminate remediation wastes effort; VulStamp is explicit about this framing \cite{shen2025vulstamp}. The early-exploitability study restricts itself to disclosure-time information so the prediction is usable immediately after a CVE is created \cite{iannone2024early}. Risk assessment adds consequences a score omits: the Model Context Protocol framework positions itself as a dynamic alternative to static CVSS-based assessment \cite{dhok2026mcp}, an explainable risk-management study ties exploitation to insurability decisions \cite{papastergiou2026xai}, a risk-aware verification engine calibrates checking effort from a composite risk score \cite{alharby2026ravecode}, and a multi-agent framework addresses threat mitigation for AI-based systems \cite{foundjem2025multiagent}. Prioritization is the terminal class and appears less populated than assessment in this corpus: POLAR automates cyber threat prioritization for analysts \cite{tang2025polar}, and the boundary study VulnRepairEval tests whether exploit-based verification changes the ranking of remediation outcomes \cite{wang2025vulnrepaireval}. Fig.~\ref{fig:classes} maps the 23 primary studies onto the five classes (Table~\ref{tab:citation_index} indexes all 23); several contribute to more than one, and assessment is the largest class, followed by risk and prioritization.

% == Figure 2 ==
\vspace{-4pt}
% == Table IV (citation index) ==
\begin{table*}[!t]
\centering
\caption{Citation Index: All 23 Primary Studies and Their Application Classes}
\label{tab:citation_index}
\scriptsize
\renewcommand{\arraystretch}{1.0}
\begin{minipage}[t]{0.495\textwidth}
\begin{tabularx}{\linewidth}{@{}c L c L@{}}
\hline
\textbf{Ref.} & \textbf{Study (short)} & \textbf{Year} & \textbf{Class(es)}\\
\hline
\cite{ali2025rag} & RAG-GPT-SBR severity prediction for bug reports & 2025 & Assessment\\
\cite{wang2025vulnrepaireval} & VulnRepairEval exploit-based evaluation & 2025 & Assessment, Prioritization\\
\cite{pakshad2026windows} & Windows App SDK vulnerability discovery analysis & 2026 & Prediction\\
\cite{pan2025alarms} & LLM-enhanced probabilistic static-alarm modelling & 2025 & Prediction\\
\cite{isogai2024cvss} & GPT-4o-mini versus BERT for CVSS base vectors & 2024 & Assessment\\
\cite{sanvito2025autocvss} & AutoCVSS automated vulnerability scoring & 2025 & Assessment\\
\cite{ghosh2025cvellm} & CVE-LLM ontology-assisted vulnerability evaluation & 2025 & Assessment\\
\cite{miranda2025contextual} & Contextual CVSS scoring for vulnerability batches & 2025 & Assessment\\
\cite{giriraju2026cti} & Threat-intelligence summarization agent for CVEs & 2026 & Assessment, Risk\\
\cite{shen2025vulstamp} & VulStamp vulnerability assessment & 2025 & Exploitability\\
\cite{iannone2024early} & Early exploitability of just-disclosed CVEs & 2024 & Exploitability\\
\cite{dhok2026mcp} & Model Context Protocol vulnerability risk assessment & 2026 & Risk, Prioritization\\
\hline
\end{tabularx}
\end{minipage}\hfill
\begin{minipage}[t]{0.495\textwidth}
\begin{tabularx}{\linewidth}{@{}c L c L@{}}
\hline
\textbf{Ref.} & \textbf{Study (short)} & \textbf{Year} & \textbf{Class(es)}\\
\hline
\cite{papastergiou2026xai} & Explainable AI risk management for insurability & 2026 & Risk\\
\cite{alharby2026ravecode} & RAVE-Code risk-aware verification engine & 2026 & Risk, Prioritization\\
\cite{foundjem2025multiagent} & Multi-agent framework for threat mitigation & 2025 & Risk, Prioritization\\
\cite{tang2025polar} & POLAR cyber threat prioritization & 2025 & Exploitability, Risk, Prioritization\\
\cite{felkner2024iot} & IoT vulnerability and attack repository & 2024 & Prediction, Assessment, Risk\\
\cite{zheng2025bounty} & Bug-bounty report invalid-reason analysis & 2025 & Assessment\\
\cite{alam2024ctibench} & CTIBench cyber threat intelligence benchmark & 2024 & Evaluation\\
\cite{cheng2026cticonnect} & CTIConnect retrieval-augmented CTI benchmark & 2026 & Evaluation\\
\cite{xu2024contamination} & Benchmark data contamination of LLMs & 2024 & Validity\\
\cite{alhaddad2026temporal} & Temporal paradox in vulnerability prioritization & 2026 & Exploitability, Prioritization\\
\cite{seo2025autopatch} & AutoPatch multi-agent CVE patching & 2025 & Prioritization, Validity\\
\hline
\end{tabularx}
\end{minipage}
\end{table*}

% == Figure 3 ==

%% ==
\subsection{RQ2 (How They Do It): Models, Adaptation and Architecture}\label{sec:rq2}

\begin{figure}[!t]
\centering
% PLACEHOLDER FOR FIGURE 5: replace figure5.png (originally fig4_adaptation.png)
\figimage{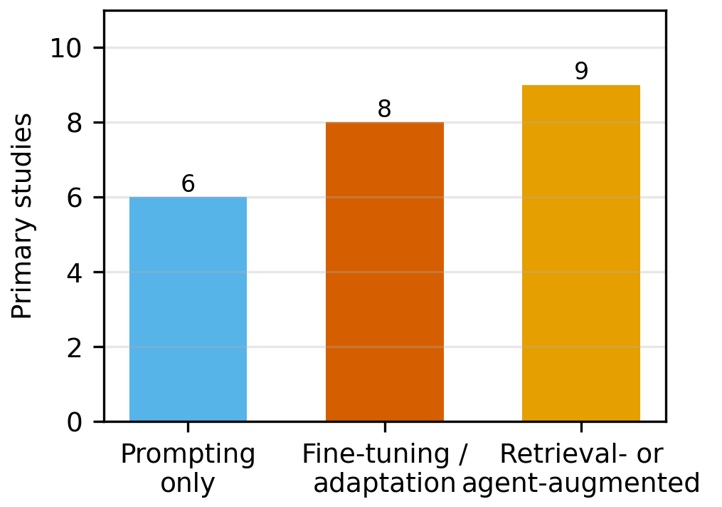}{0.78}
\caption{Model adaptation levels used across the primary studies.}
\label{fig:adaptation}
\end{figure}

The engineering of these systems is more uniform than the task taxonomy might suggest. Almost every included study begins from a general-purpose chat model and then decides how much task-specific machinery to add. Three adaptation levels recur (Fig.~\ref{fig:adaptation}): using the model as-is with a crafted prompt, adapting it with task-specific training, and wrapping it in a retrieval or multi-agent architecture that supplies external context before the model produces its answer. Studies rarely justify the choice among them against the alternatives, which is itself a finding: studies attack the same task at different levels, so reported results are not comparable in the way a controlled ablation would make them.

\subsubsection{Model Families and Proprietary versus Open-Weight Reliance}

Model choices cluster around a small number of families. Proprietary chat models dominate the assessment and prioritization studies, while open-weight and encoder-style models appear more often in prediction and classification settings; the CVSS-vector comparison is explicit about this pairing, pitting a hosted chat model against an encoder baseline on an identical task \cite{isogai2024cvss}. The split has a direct methodological consequence. Researchers address proprietary models through a hosted interface, so a study can report a prompt but not the weights, decoding configuration or training mixture, whereas open-weight models permit local inspection but require the study to publish adapters for the result to be checkable. Structured assessments of artefacts such as the Windows App SDK depend on hosted reasoning over tool output, and therefore inherit this opacity \cite{pakshad2026windows}.

\subsubsection{Prompting, Fine-Tuning and Retrieval-Augmented Pipelines}

At the prompting level, studies range from single-shot severity questions to structured prompts that request metric-by-metric CVSS reasoning or a structured weakness label. Fine-tuning appears as full supervised tuning and as parameter-efficient adaptation on vulnerability corpora, and studies usually claim it helps most on tasks with a stable label space such as severity band classification; the threat-intelligence summarization agent is an example, fine-tuned to emit severity, contextual risk and a mitigation recommendation in one pass \cite{giriraju2026cti}. Retrieval augmentation is the most active area: studies ground the model in advisory text, weakness descriptions, repository context or threat intelligence before asking for a judgement. CVE-LLM is the clearest case, retrieving ontology-grounded vulnerability evidence to support evaluation \cite{ghosh2025cvellm}, while a domain-specific repository supplies the kind of heterogeneous, semi-structured vulnerability and attack data such retrieval layers depend on \cite{felkner2024iot}. The mechanism is the same across studies even when the corpus differs: retrieve evidence, place it in context, then constrain the answer to the target label or score space.

\subsubsection{Agentic and Multi-Model Architectures}

The most recent layer decomposes the task across several model calls. In these designs one component retrieves or summarises evidence, another produces a candidate score or ranking, and a third checks or aggregates the result before the output is committed. A structure-aware contextual framework for vulnerability risk assessment is representative of this decomposition \cite{dhok2026mcp}, and a multi-agent design for threat mitigation and resilience extends it to coordinated defence tasks \cite{foundjem2025multiagent}. The motivation is auditability as much as accuracy: a decomposed pipeline leaves intermediate artefacts that an analyst can inspect, which a single prompt does not. The cost is that each added component introduces a failure mode and a configuration surface \cite{ullah2022scalability}, and the included studies report prompt or orchestration artefacts far less often than the final metric, which is why the reproducibility discussion in RQ5 treats architectural complexity as a liability as well as a capability.

% == Figure 5 ==

%% ==
\subsection{RQ3 (What They Are Evaluated On): Data, Ground Truth and Metrics}\label{sec:rq3}

Evaluation practice is where the literature is least comparable. Studies share a small pool of public ground truth but define labels, splits and success differently enough that two numbers from two papers cannot be read as the same measurement. This section inventories the recurring datasets and metrics, then isolates the two design choices that most strongly determine whether a result transfers.

\subsubsection{Datasets and Registries: CVE, NVD, CWE, CVSS, KEV and EPSS}

\begin{figure}[!t]
\centering
% PLACEHOLDER FOR FIGURE 6: replace figure6.png (originally fig5_datasets.png)
\figimage{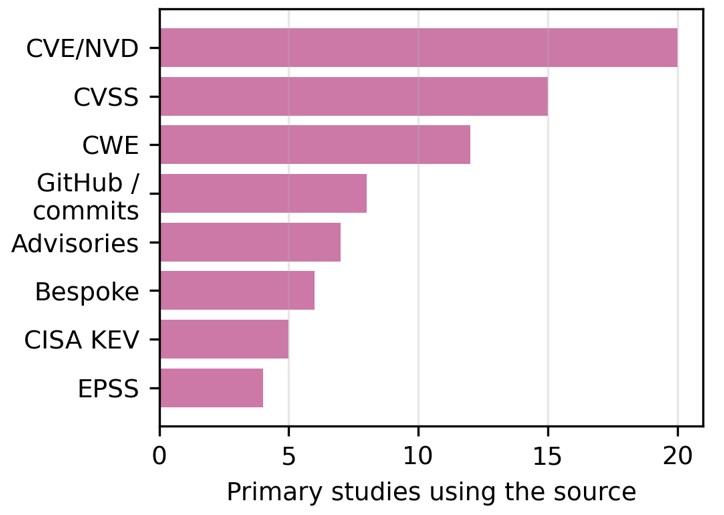}{0.78}
\caption{Ground-truth and data sources used across the corpus.}
\label{fig:datasets}
\end{figure}

The dominant ground truth is the public vulnerability record (Fig.~\ref{fig:datasets}): CVE identifiers and text, CVSS vectors and base scores, CWE mappings, and, in the exploitation and prioritization strands, CISA KEV membership and EPSS probability scores. The semantics are well defined. CVSS v4.0 combines Base, Threat, Environmental and Supplemental metric groups into a 0.0-10.0 score with explicit nomenclature, so a study targeting only base metrics is predicting a CVSS-B value and is not interchangeable with one that also models threat or environmental context \cite{miranda2025contextual}, \cite{sanvito2025autocvss}. Exploitation ground truth is sparser and more temporal: EPSS publishes a daily probability per CVE and KEV is a curated set of exploited vulnerabilities, so a ``will this be exploited'' label depends on the observation window and curation policy \cite{iannone2024early}. Not all data is registry-shaped: a domain-specific repository for IoT vulnerabilities and attack patterns shows how much relevant evidence lives outside NVD, in collections that differ in schema and completeness \cite{felkner2024iot}. A minority of studies build bespoke datasets from advisories, bug reports or repository history \cite{ali2025rag}, \cite{zheng2025bounty}, \cite{tyllis2023bigdata}.

\subsubsection{Metrics, Splits and Cross-Study Comparability}

Metrics divide along task type: classification-style severity work reports accuracy, precision, recall and F1 over severity bands; score regression reports mean absolute error and root mean squared error against CVSS values; prioritization reports ranking measures such as precision at k or the proportion of truly exploited vulnerabilities captured near the top of the queue. The choice is often reasonable in isolation but damaging in aggregate, because a study optimising band accuracy and a study minimising score error are not competing on the same objective, and neither reports the other's metric. Evaluation corpora now exist for the threat-intelligence layer these systems consume: CTIBench tests model reliability in cyber threat intelligence, and CTIConnect extends this to retrieval-augmented models over heterogeneous sources \cite{alam2024ctibench}, \cite{cheng2026cticonnect}. These benchmarks advance the field because they test the evidence-processing step directly, yet they measure the model rather than the vulnerability decision, so they do not make prioritization results comparable across studies.

\subsubsection{Benchmark Contamination and Temporal Validity}

Two validity problems recur, and researchers now study them explicitly. The first is benchmark contamination: because developers train language models on large web and code corpora, models may memorise evaluation items predating the training cutoff rather than infer them, and dedicated surveys catalogue the mechanisms and detection strategies for this in the general case \cite{xu2024contamination}. The second is temporal sensitivity. A prioritization study that varies the knowledge cutoff shows performance is not uniform across time and can even be higher on post-cutoff items, contradicting a simple memorisation story and pointing to distributional differences in the record \cite{alhaddad2026temporal}. Exploit-based verification reinforces the point: when studies tighten the criterion from superficial textual overlap to actual exploit confirmation, reported capability falls, implying earlier numbers on the same task were partly an artefact of the validation method \cite{wang2025vulnrepaireval}, \cite{seo2025autopatch}. A single random split therefore cannot establish prospective validity.

% == Figure 6 ==

% == Table V (data-extraction schema) ==
\begin{table}[!t]
\centering
\caption{Data-Extraction Schema and Coverage Across the 23 Primary Studies}
\label{tab:extraction}
\scriptsize\renewcommand{\arraystretch}{1.0}
\begin{tabularx}{\columnwidth}{l L c}
\hline
\textbf{Category} & \textbf{Extracted field} & \textbf{Coverage}\\
\hline
Bibliographic & Year, venue, publisher, DOI & 23/23\\
Application & Prediction / severity / exploitability / risk / prioritization & 23/23\\
LLM & GPT, Llama, CodeLlama, Gemini, encoder baselines & 22/23\\
Model type & Proprietary vs open-weight & 21/23\\
Approach & Zero/few-shot, fine-tuning, RAG, agent, ensemble & 23/23\\
Input & CVE text, code, CWE, CVSS vector, advisories, commits & 23/23\\
Dataset & NVD, CVE, CISA KEV, GitHub, advisory, bespoke & 21/23\\
Output & Class, CVSS score, exploit probability, risk level, ranking & 23/23\\
Baseline & CVSS, EPSS, ML/DL, static analysis, human expert & 18/23\\
Metrics & Accuracy, P/R/F1, AUC, MAE/RMSE, Precision@K & 20/23\\
Explainability & Yes/No plus technique & 12/23\\
Validation & Benchmark / real-world / industrial & 16/23\\
Reproducibility & Code, data or prompts public & 7/23\\
\hline
\end{tabularx}
\end{table}

%% ==
\subsection{RQ4 (How Well They Work): Reported Effectiveness Against Baselines}\label{sec:rq4}

Reported effectiveness is high but unevenly distributed across the taxonomy, and the distribution is itself informative. Studies that predict a bounded label space from descriptive text report the strongest results; studies that must forecast adversary behaviour or produce a defensible ranking report the weakest or the most qualified results. Crucially, the optimistic numbers concentrate in evaluation designs that the previous section identified as vulnerable to memorisation, so we cannot report the two facts separately without misleading the reader. We therefore state the performance picture below (Table~\ref{tab:performance}) together with the design that produced it. All performance figures are author-reported and were not re-run; because datasets, splits, metrics and baselines differ across studies, we do not pool them, and Table~\ref{tab:performance} shows direction of effect rather than magnitude.

\subsubsection{Severity and CVSS Scoring Performance}

Severity work is the most mature \cite{abdulsatar2025cyberwise}. Comparative studies report that general chatbot models can predict CVSS base vectors at rates well above trivial baselines \cite{isogai2024cvss}, and a systematic assessment of automated scoring across models and scoring variants reports that agreement improves when studies decompose the task metric by metric rather than answer it as a single number \cite{sanvito2025autocvss}. Ontology-assisted evaluation pushes in the same direction: supplying structured weakness context raises agreement with held-out annotations \cite{ghosh2025cvellm}, and treating related vulnerabilities as a batch rather than isolated items reduces inconsistency in the assigned scores \cite{miranda2025contextual}. Retrieval augmentation helps most where the input is under-specified, which is why a retrieval-augmented pipeline improves bug-report severity prediction over a prompting-only baseline \cite{ali2025rag}, and why a fine-tuned agent can emit a severity label, contextual risk reading and mitigation recommendation in one pass \cite{giriraju2026cti}.

\subsubsection{Exploitability and Prioritization Performance}

Exploitability and prioritization results are weaker and more design-dependent. Exploitability prediction has a strong statistical baseline in EPSS, and the disclosure-time study is careful to use only information available at CVE creation, which is what makes its early prediction operationally meaningful but also harder to score well \cite{iannone2024early}. VulStamp attacks the same problem from the triage side, predicting exploitability to avoid spending remediation effort on low-impact flaws \cite{shen2025vulstamp}. Prioritization inherits this difficulty: a ranking is only as good as the exploitation signal driving it, so studies report how many exploited or high-priority items appear near the top of the queue rather than classification accuracy. POLAR automates threat prioritization for analysts and reports gains over manual triage \cite{tang2025polar}, while a risk-aware verification engine shows that calibrating analysis effort from a composite risk score changes how much checking high-risk code receives \cite{alharby2026ravecode}. Alarm-level prioritization for static analysis is the closest analogue in the code domain, ordering findings so reviewers reach the valuable ones first \cite{pan2025alarms}.

\subsubsection{Performance Against CVSS, EPSS, ML and Human Baselines}

Placed side by side, the baselines tell a consistent story. Against classical machine learning on identical features, studies report that language models win on text-dominated tasks and tie on tabular tasks whose feature set already encodes the answer. Against CVSS itself, the fairest reading is agreement rather than improvement: predicting a CVSS vector is only valuable if the prediction approaches the published value, so high reported accuracy mostly shows the metric structure is learnable, not that the model judges severity better than a human analyst. Against EPSS the comparison is harder because EPSS is a calibrated probability with a public operating point, and few included studies report calibration rather than discrimination. Against human experts the evidence is thinnest, and no included study reports a controlled head-to-head in which analyst decisions are the outcome. The one general caveat applying to all four comparisons is temporal: when researchers vary the knowledge cutoff explicitly, the reported advantage is not stable, and in the prioritization setting the direction of the cutoff effect is not even monotonic \cite{alhaddad2026temporal}.

% == Table VI ==
\begin{table*}[!t]
\centering
\caption{Reported Performance and Baselines by Task Family}
\label{tab:performance}
\scriptsize\renewcommand{\arraystretch}{1.0}
\begin{tabularx}{\textwidth}{L L L L}
\hline
\textbf{Task family} & \textbf{Metric family} & \textbf{Typical baseline} & \textbf{Reported outcome}\\
\hline
Severity / CVSS scoring & MAE, RMSE, vector accuracy & Encoder or tabular ML & Matches or exceeds baseline\\
Severity band classification & Accuracy, F1 & Encoder or tabular ML & High, label-space dependent\\
CWE classification & Accuracy, F1 & Encoder classifier & Competitive; long-tail weak\\
Exploitability prediction & AUC, PR-AUC, calibration & EPSS & Competitive but not decisive\\
Known-exploited prediction & Precision, recall & KEV membership heuristics & Mixed; window sensitive\\
Risk assessment & Qualitative, rank correlation & Expert rubric & Promising, few quant.\ results\\
Prioritization / ranking & Precision@K, recall@K & CVSS-only ordering & Improved with context signals\\
All families, cutoff varied & Same as above & Same as above & Not stable across periods\\
\hline
\end{tabularx}
\end{table*}

%% ==
\subsection{RQ5 (What Blocks Deployment): Validity, Reproducibility and Operations}\label{sec:rq5}

Accuracy alone does not explain the gap between reported results and operational use. Five barriers recur, and they interact: a system can be accurate on a leaked benchmark, offer explanations nobody has tested, be hard to reproduce, be costly or privacy-limited to run, and generalise poorly, and each defect makes the others harder to detect.

\subsubsection{Hallucination, Explainability and Analyst Trust}

Language models can produce a confident severity judgement or a plausible exploitability rationale that lacks grounding in the advisory or code they received, and the risk is highest where users request a numeric score with no retrieved evidence to anchor it. CTIBench's authors built it precisely because reliability, accuracy and hallucination remain open concerns in the threat-intelligence tasks these systems depend on \cite{alam2024ctibench}, and CTIConnect shows that grounding a model in heterogeneous sources is a partial rather than complete remedy \cite{cheng2026cticonnect}. Studies respond by requesting structured reasoning, retrieving evidence before answering, or decomposing the task so an intermediate step can be checked, and several report that added structure reduces output inconsistency \cite{ghosh2025cvellm}, \cite{shen2025vulstamp}. Some studies also treat explainability as a deliverable in its own right, as in the explainable risk-management study where the rationale matters for insurability decisions \cite{papastergiou2026xai}. What the corpus lacks is a controlled test of whether those explanations change what an analyst does: no included primary study reports a human-subject evaluation in which triage decisions or trust calibration are the measured outcome, so we record the claim that LLM explanations are usable in practice as an audited shortfall rather than a finding.

\subsubsection{Temporal Leakage, Contamination and Generalizability}

The validity barrier described under RQ3 has a deployment consequence: a model that scores well by recognising evaluation items it has already seen will not reproduce that performance on tomorrow's advisories, and the failure appears after adoption rather than before. The general language-model literature now studies contamination systematically \cite{xu2024contamination}, and the domain-specific evidence shows performance varies with the knowledge cutoff in ways a random split cannot expose \cite{alhaddad2026temporal}. Model knowledge also expires against the threat it is meant to track \cite{sabir2021exfiltration,ullah2022adaptive}, which motivates an approach that avoids frequent fine-tuning on new CVE sets \cite{seo2025autopatch}. Generalizability compounds the problem \cite{ahmad2023c3i}: registries, exploit catalogues and advisory formats evolve, and a repository study shows how heterogeneous the underlying data already is across domains such as IoT \cite{felkner2024iot}. A second limit is scope: scoring a generic CVE and scoring it as it affects a specific deployed asset are different judgements \cite{jayalath2024microservice}, and developers train most models on the generic form.

\subsubsection{Cost, Privacy and Reproducibility of Published Pipelines}

The final barrier is operational. Pipelines relying on hosted proprietary models must transmit advisory text, and often code or repository context, to a third party, a governance obstacle in regulated environments; local open-weight models avoid that exposure but need compute many teams cannot provision \cite{ullah2024hybrid}, and the trade-off is rarely quantified in the studies we reviewed. Validation quality is a related weakness: exploit-based verification showed that superficially validated gains do not survive a stricter criterion, exactly the failure mode a deployment would expose \cite{wang2025vulnrepaireval}, \cite{seo2025autopatch}. Reproducibility is the weakest reported dimension overall; studies sometimes publish prompts, but they usually omit adapters, orchestration code, random seeds and cost figures, and the orchestration-heavy designs are the hardest to rebuild \cite{foundjem2025multiagent}, \cite{dhok2026mcp}. A reader can therefore compare headline numbers but cannot rerun the experiment that produced them.

% == Table VII ==
\begin{table*}[!t]
\centering
\caption{Threats to Validity and Mitigations}
\label{tab:threats}
\scriptsize\renewcommand{\arraystretch}{1.0}
\begin{tabularx}{\textwidth}{L l L L}
\hline
\textbf{Threat} & \textbf{Type} & \textbf{Mitigation applied} & \textbf{Residual risk}\\
\hline
Six named databases not directly reachable & External & OpenAlex venue indexing, arXiv, DOI full text, official registries & Venue coverage bias\\
Semantic Scholar rate limiting & External & Reran affected cells on OpenAlex/arXiv & Minor cell-level gaps\\
Abstract-level evidence for some venues & Internal & Bounded all claims; flagged in extraction notes & Understated QA5/QA6\\
Single-reviewer screening & Internal & Deterministic criteria; reproducible funnel & No Cohen's kappa\\
Author-reported performance not reproduced & Internal & We qualified results as author-reported throughout & Effect sizes uncertain\\
Benchmark contamination & Construct & We treated it as an analytical theme & Cannot retro-correct studies\\
Registry facts time-sensitive & External & We cited official sources & Point-in-time values\\
\hline
\end{tabularx}
\end{table*}

%% ====================================================================
\section{Conclusion}\label{sec:conclusion}

Large language models have moved into the vulnerability-management stages that follow detection, but the transition is uneven and the evidence supporting it is thinner than the volume of published work suggests. The taxonomy established in this review separates five task classes, and the evidence base is densest where the label space is tightest and sparsest where the operational decision is hardest: severity and CVSS assessment are well populated, exploitability forecasting depends on a strong statistical baseline, and prioritization, the class that would actually change remediation order, has fewer studies than assessment and more qualified evidence. The methods used to reach these results are convergent despite the diversity of tasks, which means the field's methodological risk is shared rather than spread: reliance on hosted models, random-split evaluation and incomplete reporting recur across studies that otherwise differ.

Four recommendations follow. First, evaluation should adopt time-ordered splits by default, training on earlier records and testing on later ones, and reporting the knowledge cutoff explicitly, because random splits cannot distinguish inference from memorisation in a domain where the benchmark predates the model. Second, studies should publish the artefacts that make a pipeline checkable, namely prompts, adapters, orchestration code, seeds and inference cost, since reproducibility is the weakest dimension of the corpus. Third, prioritization work should report ranking-aware and calibration-aware metrics against operational baselines rather than classification accuracy alone. Fourth, authors should scope claims to the class of judgement their model actually makes, because scoring a generic CVE and ranking it for a specific asset are different tasks with different validity requirements.

Three validation paths would settle the questions this review could not. An analyst study should measure whether LLM-generated explanations change triage decisions and trust calibration, closing the shortfall recorded here. A public, time-split benchmark with a fixed observation window for exploitation labels would make exploitability and prioritization results comparable across studies for the first time. Finally, a cost-and-privacy-aware comparison of open-weight and hosted pipelines on identical tasks and splits would convert the current anecdotal trade-offs into evidence a security team can act on. Until those exist, the defensible position is that these systems should support human analysers rather than replace their judgement, and that readers should read their reported accuracy as an upper bound conditioned on the evaluation design that produced it.

%% ====================================================================
\bibliographystyle{IEEEtran}
\bibliography{references}

\end{document}